\documentclass[a4paper,fleqn,usenatbib]{mnras}
\usepackage[T1]{fontenc}
\usepackage{ae,aecompl}
\usepackage{epsfig}
\usepackage{pslatex}
\usepackage{amsmath}
\usepackage{fleqn}
\usepackage{breqn}

\defcitealias{Zuc2015}{Paper~I}

\title[Improved Non-Parametric Periodicity Detection]{
Detection of Periodicity Based on Independence Tests -- II. Improved Serial Independence Measure}
\author[Shay Zucker]{Shay Zucker\thanks{E-mail:
shayz@post.tau.ac.il}\\ Dept. of Geosciences, Raymond and Beverly
Sackler Faculty of Exact Sciences, Tel-Aviv University, Tel Aviv
6997801, Israel}

\begin{document}
\label{firstpage}
\pagerange{\pageref{firstpage}--\pageref{lastpage}}
\maketitle

\begin{abstract}	

We introduce an improvement to a periodicity metric we have introduced
in a previous paper. We improve on the Hoeffding-test periodicity
metric, using the Blum-Kiefer-Rosenblatt (BKR) test. Besides a
consistent improvement over the Hoeffding-test approach, the BKR
approach turns out to perform superbly when applied to very short time
series of sawtooth-like shapes. The expected astronomical implications
are much more detections of RR-Lyrae stars and Cepheids in sparse
photometric databases, and of eccentric Keplerian radial-velocity (RV)
curves, such as those of exoplanets in RV surveys.

\end{abstract}


\section{Introduction}

In a recent paper \citep[][; hereafter \citetalias{Zuc2015}]{Zuc2015}
we have introduced a new non-parametric approach to the detection of
periodicities in sparse datasets. The new approach follows the logic
of string-length techniques \citep[e.g.][]{LafKin1965,Cla2002} and
quantifies the dependence between consecutive phase-folded samples,
for every trial period. In \citetalias{Zuc2015} we have shown that
usually the classical string-length techniques effectively test solely
for linear dependence between consecutive samples. On the other hand,
the Hoeffding-test approach we have presented there, tests for general
dependencies, not necessarily linear. \citetalias{Zuc2015} showed that
for two kinds of periodic signals (sawtooth signal and eccentric
spectroscopic binary (SB) radial-velocity (RV) curve), our proposed
new approach performed better than the conventional techniques.

Inspired by the successful simulations presented in
\citetalias{Zuc2015}, we embarked on a wider study to come up with new
and improved periodicity metrics, similarly based on dependence
measures. The current Letter represents a first step in that
direction, a step we have already alluded to in the Discussion of
\citetalias{Zuc2015}.

Section \ref{BKR} describes the details of the modification we propose
to the Hoeffding test, Section \ref{performance} presents the
simulations we have performed in order to test its
performance, and in Section \ref{Hipparcos} we show a test of
our new metric on real-life data.  In Section \ref{discussion} we
discuss our findings.

\section{Blum-Kiefer-Rosenblatt test}
\label{BKR}

The best performing method in \citetalias{Zuc2015} was based on the
Hoeffding test. Wassily Hoeffding first proposed it in 1948
\citep{Hoe1948} as a test of independence between two random
variables. Essentially it estimates a measure of deviation of the
joint empirical distribution function from a distribution that assumes
independence. The measure of deviation that Hoeffding used was the
so-called Cram\'er--von Mises criterion for distance between
distributions \citep{Cra1928,von1928}.

Let us denote by $G_1$ and $G_2$ the cumulative distribution functions
of the two random variables, and by $G_{12}$ their joint cumulative
distribution function. Then, independence of the two variables would
mean $G_{12} = G_1 G_2$.  Applying Cram\'er--von Mises criterion for
distance between distributions, Hoeffding defined his namesake
statistic $D$ by: 
\begin{equation}
D = \int (G_{12} - G_1 G_2)^2\,dG_{12} \ .
\end{equation}

In estimating $D$ using the emprirical data, we use the empirical
distribution functions, determined only by the observed values. This
is somewhat reminiscent of the Kolmogorov--Smirnov philosophy, which
is popular among astronomers \citep[e.g.][]{BabFei2006}.  The above
definition eventually results in the formulae presented in
\citetalias{Zuc2015}.

\citet*{Bluetal1961} introduced a new version of the Hoeffding
test. They showed that the two tests were equivalent in large
samples, but the new test was easier to compute, and also more
naturally amenable to generalization to more than two variables. The
fundamental definition of their statistic is:
\begin{equation}
B = \int (G_{12} - G_1 G_2)^2\,dG_1\,dG_2 \ .
\end{equation}
While the difference seems to be minute and maybe even insignificant,
the change in the resulting computing formula is not negligible.

Following \citetalias{Zuc2015}, let us denote the phase-folded data by
$x_i$~($i=1,...,N$) where the index $i$ reflects the order after the
phase folding. In order to make sure our calculations are not affected
by the arbitrary zero-phase choice, we also define $x_{N+1} \equiv
x_1$.  Let us further denote by $R_i$ the phase-folded rank values,
such that $R_i=1$ means that $x_i$ is the smallest value.  Let us also
define the 'bivariate rank' $c_i$, as the number of pairs
$(x_j,x_{j+1})$ for which both $x_j \leq x_i$ and $x_{j+1} \leq
x_{i+1}$. Note that the existence of the pair $(x_N,x_{N+1}) =
(x_N,x_1)$ accounts for the cyclic wraparound and renders the whole
procedure independent of the arbitrary choice of phase.

Now we can define our dependence measure by the formula:
\begin{equation}
B = N^{-4} \sum_{i=1}^{N} (Nc_i-R_i R_{i+1})^2 \ ,
\end{equation}
which can easily be derived from eq.\ (5.2) in \cite{Bluetal1961}.
The above expression is clearly much simpler than the parallel
expressions in eqs.\ 9--12 in \citetalias{Zuc2015}.

\section{Performance testing}
\label{performance}

Following the same path as in \citetalias{Zuc2015}, we performed
simulations in which we randomly drew a sparse set of sampling times,
from a total baseline spanning $1000$ time units ('days'). Then we
used those times to sample some periodic function with a period of two
days, and added white Gaussian noise with a prescribed signal-to-noise
ratio (SNR). We tested the same set of six periodic functions we
tested in \citetalias{Zuc2015}: sinusoidal, almost sinusoidal,
sawtooth, pulse wave, eclipsing-binary light curve and eccentric SB RV
curve.

Unlike the approach we followed in \citetalias{Zuc2015}, we have
decided this time to use a much simpler and intuitive performance
measure: in each tested configuration of signal shape, $N$ and SNR, we
simply counted the number of simulations in which the best score was
attained exactly in the correct known period, thus obtainaing the
'detection fraction'. We calculated the periodicity metrics for a
frequency grid that spanned the range
$10^{-4}$--$1\,\mathrm{day}^{-1}$, with steps of
$10^{-4}\,\mathrm{day}^{-1}$. Thus, counting the fraction of cases
with the correct period actualy meant a frequency error which was
smaller than $10^{-4}\,\mathrm{day}^{-1}$.

We compared the performance of this Blum-Kiefer-Rosenblatt-test (BKR)
periodicity metric to the Hoeffding-test metric we had introduced in
\citetalias{Zuc2015}, and also to the two 'traditional' techniques of
Generalized Lomb-Scargle \citep{Lom1976,Sca1982,ZecKur2009} and
von-Neumann ratio \citep{vonetal1941} as a representative of the
string-length techniques \citep{Cla2002,LafKin1965}.

\begin{figure}
\includegraphics[width=0.5\textwidth]{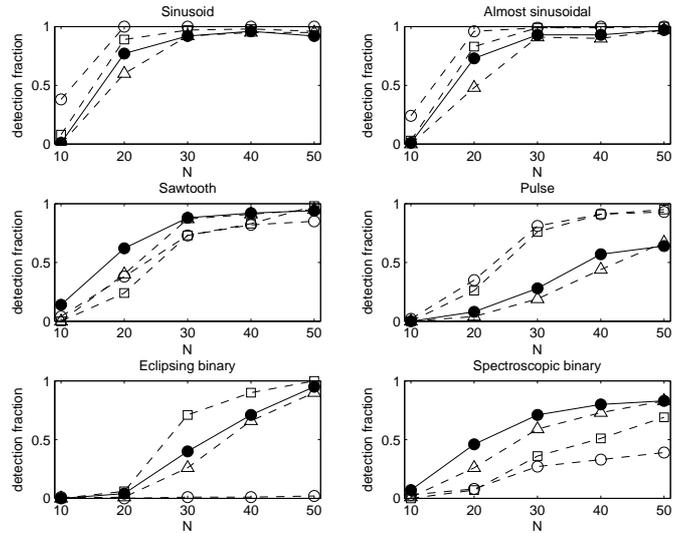}
\caption{Detection fraction as a function of sample size for time
series with SNR of $3$. The detection fraction is estimated based on
$100$ simulated light curves. Legend: empty circles, dashed line --
Generalized Lomb--Scargle periodogram; empty squares, dashed line --
von-Neumann Ratio; empty upward-pointing triangles, dashed line --
Hoeffding test; filled circles, solid line -- Blum-Kiefer-Rosenblatt
test.}
\label{SNR_3}
\end{figure}

Fig.\ \ref{SNR_3} examines the dependence of the detection performance
on the number of samples in the time series, for low-SNR. We held the
SNR fixed at $3$ while we varied the sample size $N$. The main feature
apparent from examining the Figure is that the new BKR approach
constitutes an improvement over the Hoeffding-test approach we
introduced in \citetalias{Zuc2015}. In cases where the traditional
techniques performed better (e.g., pulse-wave signal shape), they
usually performed also better than BKR, and vice versa.

\begin{figure}
\includegraphics[width=0.5\textwidth]{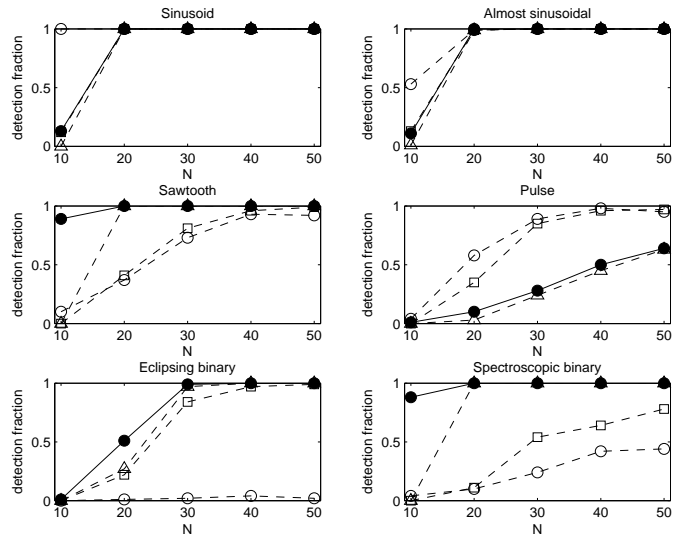}
\caption{Detection fraction as a function of sample size for time
series with SNR of $100$. The detection fraction is estimated based on
$100$ simulated light curves. Legend: empty circles, dashed line --
Generalized Lomb--Scargle periodogram; empty squares, dashed line --
von-Neumann Ratio; empty upward-pointing triangles, dashed line --
Hoeffding test; filled circles, solid line -- Blum-Kiefer-Rosenblatt
test.}
\label{SNR_100}
\end{figure}

Fig.\ \ref{SNR_100} presents the same test for the case of a high
SNR. We fixed the SNR at a value of $100$ and repeated the same
exercise. In this situation the BKR was again improving over the
Hoeffding-test approach, which also meant it performed significantly
better than the traditional approaches in the cases of sawtooth-shaped
signal and eccentric SB RV curve).  However, specifically in those two
cases, another trend emerged: the BKR method seemed to have an
exceedingly better performance for datasets with very few samples
($N=10$).

\begin{figure}
\includegraphics[width=0.5\textwidth]{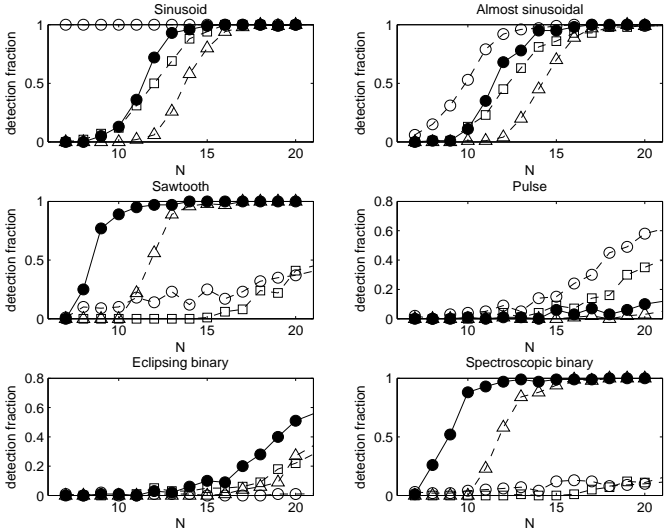}
\caption{Detection fraction as a function of sample size for time
series with SNR of $100$, focusing on small $N$. The detection
fraction is estimated based on $100$ simulated light curves. Legend:
empty circles, dashed line -- Generalized Lomb--Scargle periodogram;
empty squares, dashed line -- von-Neumann Ratio; empty upward-pointing
triangles, dashed line -- Hoeffding test; filled circles, solid line
-- Blum-Kiefer-Rosenblatt test.}
\label{SMALLN}
\end{figure}

In order to make sure this result was not spurious or even erroneous,
we have decided to examine this regime more closely. We repeated the
simulations with $\mathrm{SNR}=100$ for a finer grid of $N$, namely,
for all $N$ from $7$ to $20$. Fig.\ \ref{SMALLN} , which focuses on
this range, shows the very convincing result of a gradual increase of
the performance in most cases. Specifically in the cases of the
sawtooth signal and the eccentric SB RV curve, The pace of that
increase for the BKR method is much faster than that of the other
techniques, including the Hoeffding-test approach. It seemed that
already when there were $9$~(!) samples, the BKR approach had very
good chances to detect the periodicity.

\begin{figure}
\includegraphics[width=0.5\textwidth]{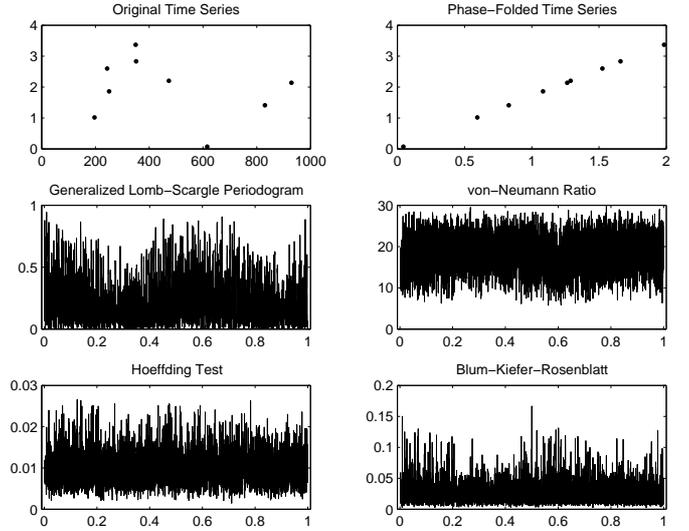}
\caption{An example of the results of applying all the examined
periodicity detection methods to a sawtooth simulated time-series,
with $9$ samples, and a SNR of $100$.  The upper two panels show the
time series and its phase-folded version, using the correct
period. The other panels show the periodicity metrics calculated for
this time series, with self explanatory titles. Note the poor
performance of the first three periodicity metrics, compared to the
detection by the BKR technique.}
\label{N9_1}
\end{figure}

\begin{figure}
\includegraphics[width=0.5\textwidth]{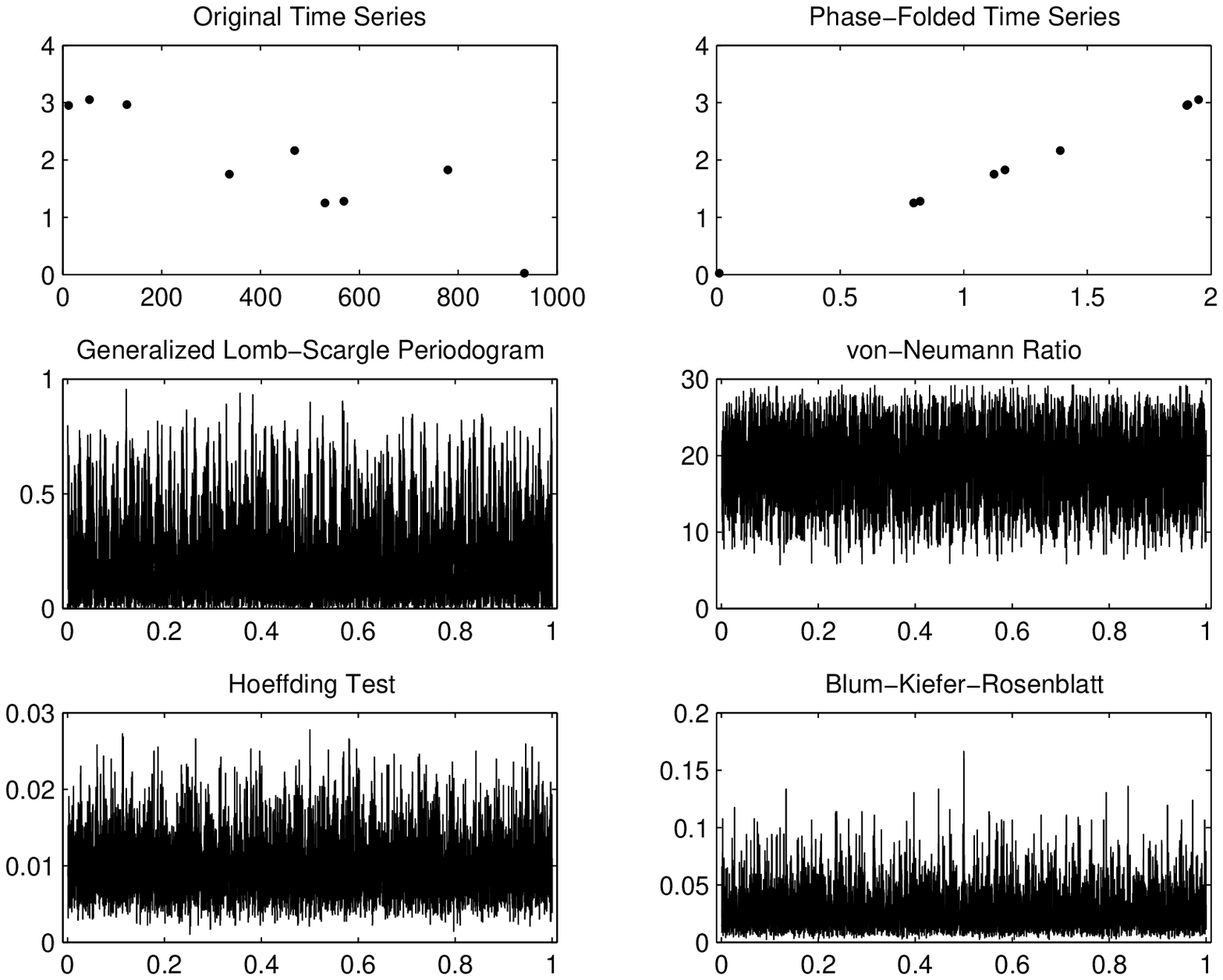}
\caption{Another example of the results of applying all the examined
periodicity detection methods to a sawtooth simulated time-series,
with $9$ samples, and a SNR of $100$.  The upper two panels show the
time series and its phase-folded version, using the correct
period. The other panels show the periodicity metrics calculated for
this time series, with self explanatory titles. Note the poor
performance of the first three periodicity metrics, compared to the
detection by the BKR technique.}
\label{N9_2}
\end{figure}

\begin{figure}
\includegraphics[width=0.5\textwidth]{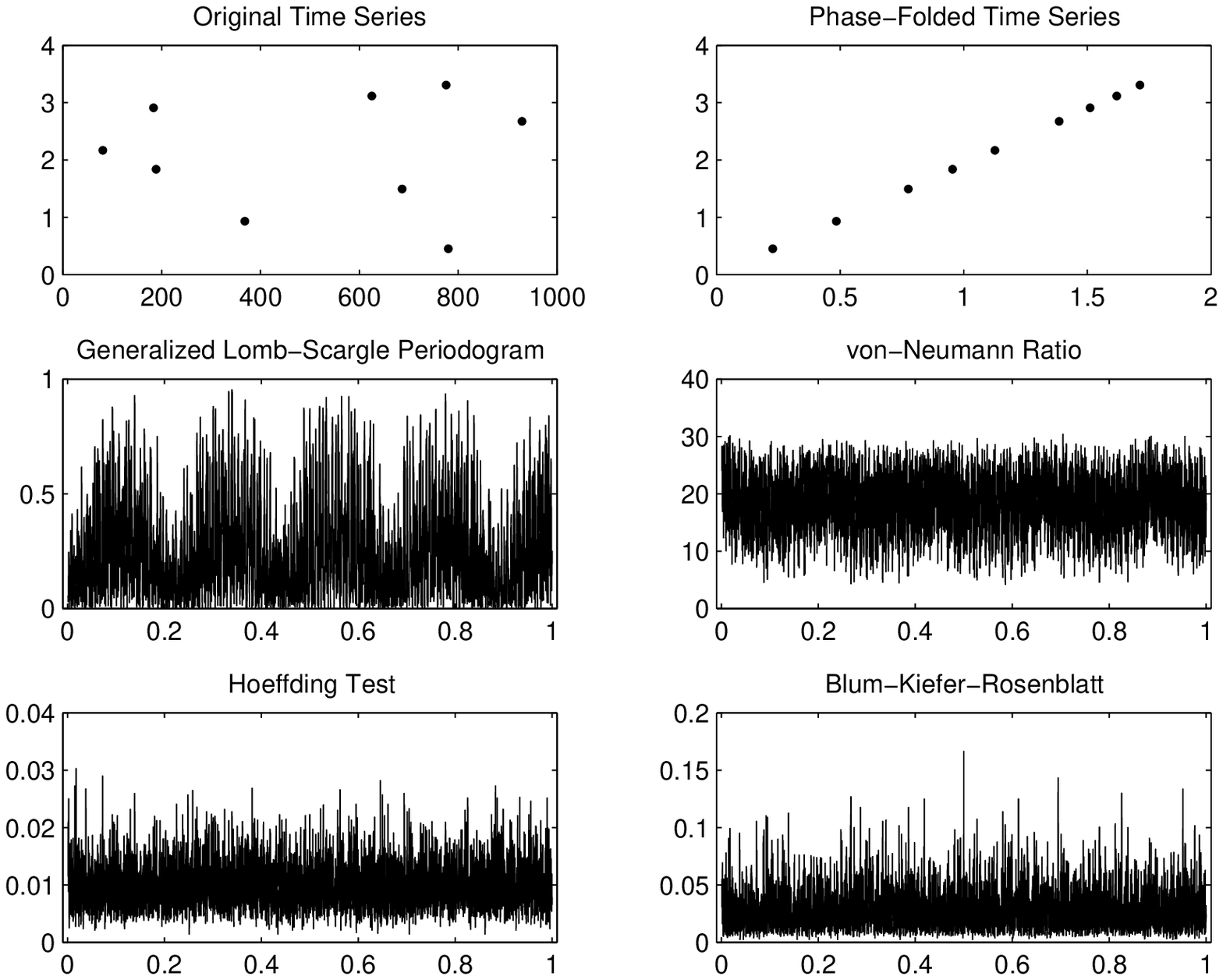}
\caption{Another example of the results of applying all the examined
periodicity detection methods to a sawtooth simulated time-series,
with $9$ samples, and a SNR of $100$.  The upper two panels show the
time series and its phase-folded version, using the correct
period. The other panels show the periodicity metrics calculated for
this time series, with self explanatory titles. Note the poor
performance of the first three periodicity metrics, compared to the
detection by the BKR technique.}
\label{N9_3}
\end{figure}

Figs. \ref{N9_1},\ref{N9_2} and \ref{N9_3} show selected concrete
examples of cases with a sawtooth signal shape and only $9$ samples,
in which the performance of BKR was definitely superior over the three
other techniques we tested. Those cases were indeed the majority of
the simulated cases.

\section{Real-life test}
\label{Hipparcos}

We set out to test our new technique in a real-life situation that
could potentially emphasize its advantages and allow them to
materialize. To this end we chose to apply it to short
\textit{Hipparcos} lightcurves. Our sample consisted of
\textit{Hipparcos} targets with at most $30$ samples in their
\textit{Hipparcos} Epoch Photometry Annex entry. We considered only
samples which were fully accepted by at least one of
\textit{Hipparcos}' FAST and NDAC consortia \citep{ESA1997}. In total
there were $51$ lightcurves meeting those criteria.

We scanned the selected lightcurves for periodicity using the BKR
test, on a frequency grid ranging from $10^{-4}$ to
$12\,\mathrm{d}^{-1}$ (following \citet{KoeEye2002}), with steps of
$10^{-4}\,\mathrm{d}^{-1}$. We searched for targets whose peak in the
BKR function was significantly prominent. We therefore used the SDE
(Signal Detection Efficiency) statistic originally used by
\citet{Alcetal2000} and \citet{Kovetal2002}. We chose to single out
targets whose SDE statistic was higher than $15$. Only one target
passed this hurdle -- HIP~101453 (also known as CH~Aql).

\begin{figure}
\includegraphics[width=0.5\textwidth]{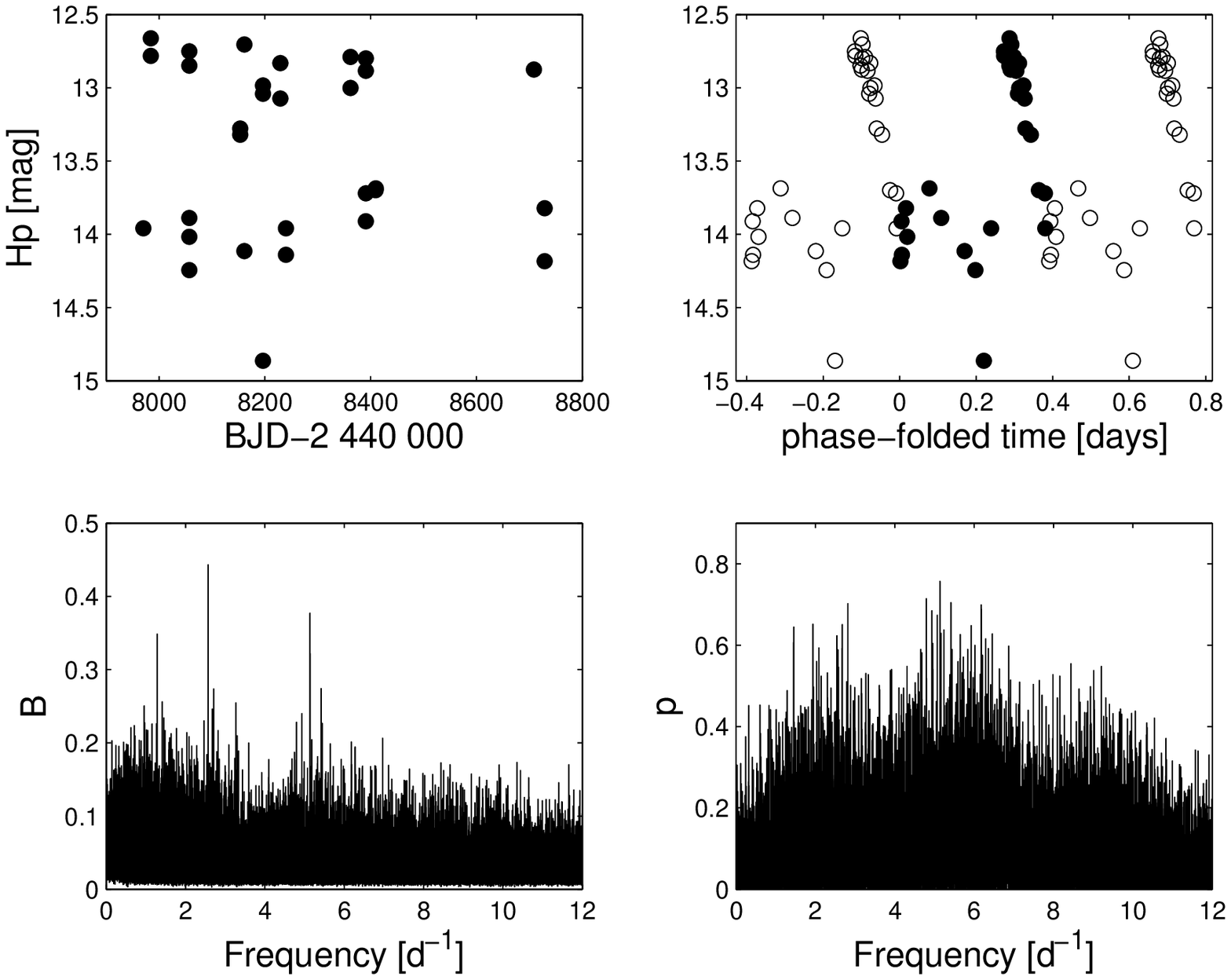}
\caption{Detecting the periodicity in the \textit{Hipparcos}
lightcurve of HIP~101453.  Upper left: the original lightcurve.  Upper
right: The lightcurve phase-folded using a period of
$0.3891656$\,d. The empty circles represent copies of the original
dataset shifted backwards and forwards by one period in order to
better visualize the periodicity. Lower left: The BKR periodicity
metric. Note the sharp peak at the known frequency. Lower right:
Generalized Lomb-Scargle periodogram. Note the absence of any
significant peak.}
\label{hip101453}
\end{figure}

Fig.\ \ref{hip101453} shows the lightcurve of HIP~101453, as well as
its phase folding using the resulting period, the BKR function and the
GLS periodogram we calculated for comparison. The prominence of the
BKR peak at a frequency of $2.5696\,\mathrm{d}^{-1}$ is evident, and
indeed the corresponding SDE value is $17.5$. The phase-folded
lightcurve demonstrates an obvious periodicity. On the other hand, The
GLS periodogram shows no hint of a statistically significant
periodicity detection.

To complete the picture, this star is a known RR-Lyrae star, with a
period of $0.38918702$\,d \citep{Sametal2013}, which is consistent
with our result -- $0.3891656$\,d.  The \textit{Hipparcos} catalogue
quotes the known period as taken from literature, and adds a comment
about the scarcity of the data, which casts doubt about the nature of
the periodicity. Therefore, the main catalogue does not quote any
period. We show here that by using our new BKR approach, we can assign
a high degree of credibility to the periodicity, even with the very
scarce \textit{Hipparcos} data.

\section{Discussion}
\label{discussion}

This Letter presents an improvement of the serial Hoeffding-test
periodicity metric we have presented in \citetalias{Zuc2015}, based on
the Blum-Kiefer-Rosenblatt modification of the original Hoeffding
test. This new periodicity metric consistently outperforms the
Hoeffding-test metric. On top of it, it seems to perform superbly
better in sawtooth-like signal shapes, when the number of samples is
very small and the SNR is high. This result is in line with the
statement of \citet{Bluetal1961}, that their test is asymptotically
equivalent to the Hoeffding test for large samples.

This advantage might prove very important for radial-velocity surveys
searching for exoplanets. The detection of high-eccentricity Keplerian
RV curves is notoriously difficult when based on a small number of
samples \citep{Cum2004}. Another situation in which this feature will
prove valuable is the detection of RR-Lyrae stars and Cepheids (whose
signal shapes are also essentially sawtooth-like shapes) in sparse
datasets such as those of \textit{Hipparcos} and \textit{Gaia}
\citep{Eyeetal2012}. We have provided in Section
\ref{Hipparcos} a demonstration of this potential using
\textit{Hipparcos} data, specifically for the case of HIP~101453.

We continue our investigation of harnessing the power of
non-parametric independence measures for the purpose of detecting
periodicity in extreme circumstances of either poor SNR, small sample
sizes, or non-sinusoidal signal shapes. In the meanwhile we also apply
our newly developed approaches to existing datasets, both for the sake
of testing, but also for detecting previously missed periodic
variables. To promote further research and testing of this periodicity
metric by the community, we make it available online, in the form of a
\textsc{MATLAB} function\footnote{The URL for downloading a
\textsc{MATLAB} code to calculate the BKR periodicity metric is
http://www.tau.ac.il/\~{}shayz/BKR.m}.

\bsp	
\label{lastpage}
\end{document}